# Spectral representation for on-shell massless double box

N. I. Ussyukina

*Institute for Nuclear Physics, Moscow State University, 119899 Moscow, Russia*
E-mail address: ussyuk@theory.npi.msu.su

A method is presented for obtaining the ε-expansion
for on-shell massless scalar double-box diagram.

**1**. There are several reasons why calculation of two-loop diagrams is important. The study of two-loop diagrams is required by increasing precision of experiments. Examination of multi-jet processes requires analytical calculations of two-loop diagrams. At last, consideration of such diagrams helps us to develop and apply new methods of loop calculations.

One-loop on-shell four-point diagram in $\varphi^4$-theory has been considered in [1]. As compared with the one-loop case, the evaluation of two-loop diagrams is technically much more complicated and less investigated. Off-shell double-box problem in 4-dimensional space-time was solved in [2]. On-shell double-box problem is more complicated than the off-shell one because the appearence the infrared divergences. Our consideration will be in the framework of dimensional regularization [3]. On-shell double-box diagram has very high degree of divergences. Let $n = 4 + 2\varepsilon$ be the space-time dimension. Then on-shell double-box diagram $D(s,t,k_i^2)$ diverges as $\varepsilon^{-4}$. It is evident that the expansion up to the forth degree is a very difficult problem. On-shell problem can be solved for example by using various reduction methods (the integration-by-part method [4], uniqueness method [5]). Such approach was used in [6], but result was too combersome. For this reason it remained the problem to look for such scheme, that could lead to more visible result. Such method I'll consider in this paper. The talk is about so named Mellin expansion method. At first this method was used for multiloop calculations in [7]. With the help of this method I managed to find the visible solution for the on-shell double-box case.

The remainder of the paper is organized as follows. In Section 2 we list notations, some most useful formulas and the main idea of calculation. In Section 3 we'll carry out some spade-work before to use the Mellin expansion method. In Section 4 using the Mellin expansion method we'll obtain the spectral representation for four-point function. In Section 5 we discuss the results.

**2.** In this section we list notations, some most useful formulas and the main idea of calculation.

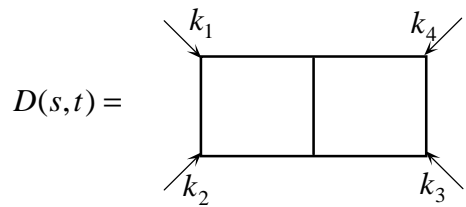

Fig.1

We use the standard notations $\quad s = (k_1 + k_2)^2, \quad t = (k_2 + k_3)^2, \quad k_i^2 = 0$.

Not to complicate formulae, we omit henceforth the factors that are powers of 2, $\pi$, i. These factors can be easily restored in the final results. Each line of diagrams carries a power-like factor that is pictured as a line with a label $\alpha$, for example,

$$\underline{\quad\alpha\quad} \Rightarrow \frac{1}{\left(p^2 - i0\right)^\alpha} \quad \text{in momentum space},$$

$$\underline{\quad\alpha\quad} \Rightarrow \frac{1}{\left(x^2 + i0\right)^\alpha} \quad \text{in coordinate space}.$$

We'll use constantly the formula for propagator Fourier-transformation from momentum to coordinate space:

$$\frac{1}{(k^2)^\alpha} \Rightarrow \frac{\Gamma(\mu - \alpha)}{\Gamma(\alpha)} \times \frac{1}{(x^2)^{\mu-\alpha}}.$$

The main idea of calculations consists in the possibility to reduce in the case $k_4^2 = k_3^2 = 0$ double-box diagram $D(s,t)$ to three-point diagram (Fig.2) $\Gamma(q_1^2, q_2^2, q_3^2)$:

$$\Gamma(q_1^2, q_2^2, q_3^2) =$$

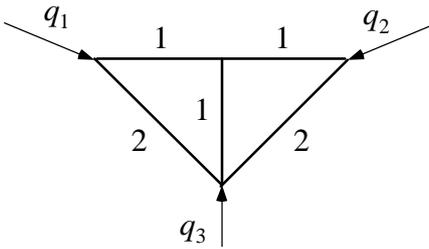

Fig.2

where

$$q_1^2 = (k_1 + k_2 x)^2, \quad q_2^2 = (k_3 + k_4 y)^2, \quad q_3^2 = t\overline{x}\overline{y}, \quad \overline{x} = 1 - x, \quad \overline{y} = 1 - y,$$

with the help of equality

$$D(s,t) = \int_0^1 dx \int_0^1 dy \, \Gamma(q_1^2, q_2^2, q_3^2). \tag{1}$$

The three-point diagram is more simple object for calculation and for expansion in $\varepsilon$ than four-point diagram and if we will know the $\varepsilon$-expansion for $\Gamma$, we'll obtain simply $\varepsilon$-expansion for $D$ from this equality.

**3**. In this section we'll carry out some preparatory work before using the Mellin-expansion method. The reason for this will be clear later. Using the uniqueness method we present the three-point diagram (here it is written in coordinate space) in the following form;

$$= \varepsilon \frac{\Gamma(1+3\varepsilon)}{\Gamma^4(1+\varepsilon)\Gamma(1-2\varepsilon)} \qquad , \qquad (2)$$

$$= 2\frac{\Gamma(2-\varepsilon)\Gamma^3(1+\varepsilon)}{\Gamma(1+2\varepsilon)} \qquad . \qquad (3)$$

We begin our calculation from the usual Feynman $\alpha$-representation for diagram of Fig.3 in coordinate representation. Then we transform this $\alpha$-representation to spectral representation for Fig. 4.

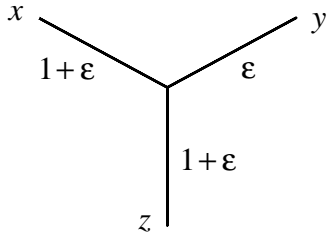
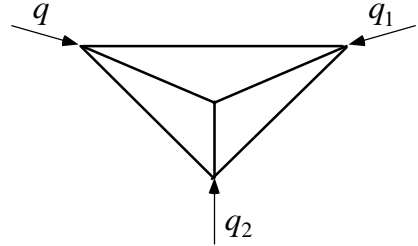

Fig. 3                          Fig.4

**4.** In this section we obtain at first spectral representation for diagram of Fig.3, then using the Mellin expansion method we'll reduce it to spectral representation for four-point function. Spectral representation for diagram of Fig.3 has the form

$$\int_0^\infty d\xi \int_0^\infty d\zeta \frac{1}{\{\xi\zeta+\xi+\zeta\}^{2-\varepsilon}\{(y-z)^2+(x-y)^2\xi+(x-z)^2\zeta\}^{2\varepsilon}}.$$

This expression can be presented in the form of two-fold Mellin integral [8]. After multiplying this Mellin expansion by the propagators that correspond to formula (3) and after transforming the result to momentum representation we obtain for two-loop three-point diagram of Fig.4 in momentum representation the following $\alpha$-representation

$$\int \prod_i d\alpha_i \, \delta(1-\Sigma_i \alpha_i) \frac{\alpha_1^{2\varepsilon}\alpha_2^{2\varepsilon}\alpha_3^{2-2\varepsilon}\varphi(\alpha_i)}{\{q^2\alpha_1\alpha_2+q_1^2\alpha_2\alpha_3+q_2^2\alpha_1\alpha_3\}}, \qquad (4)$$

where

$$\varphi(\alpha_i) = \int_1^\infty dx \, \frac{1}{x^{1-\varepsilon}} \int_0^1 dz \frac{z^{2\varepsilon-1}(1-z)^{2\varepsilon-1}}{\{\alpha_1 xz+\alpha_2 z+\alpha_3\}^{1+\varepsilon}}.$$

This representation is very similar to ordinary α-representation for one-loop three-point function, but the representation for two-loop case include weght-function $\varphi(\alpha_i)$.

After transformations that take into account equality (1) we obtain for four-point function:

$$D(s,t) \Rightarrow \int_0^1 dx \int_0^1 dy \int_0^\infty d\xi \int_0^\infty d\zeta \frac{\zeta^\varepsilon \xi^{2\varepsilon}}{\{q^2 + \xi + \zeta\}^\varepsilon \{\xi\zeta + q_1^2\xi + q_2^2\zeta\}^2} \int_0^\infty d\tau\, \tau^{2\varepsilon-1}(1+\tau)^{1-3\varepsilon}$$

$$\times \int_\zeta^\infty dz \frac{z^{\varepsilon-1}}{\{(1+\tau)q^2 + \xi + z\}^{1+\varepsilon}}.$$

At this stage we can see why the previous transformations were done. For the form (3) we can carry out integration over $y$. Then

$$D(s,t) \Rightarrow \int_0^1 dx \int_0^\infty d\xi \int_0^\infty d\zeta \int_0^\infty d\tau \frac{\xi\tau^{2\varepsilon-1}}{\{1+\tau+s\zeta\}\{1+\tau+t\bar{x}\xi\}} \int_0^\zeta dz \frac{1}{\{q^2\xi z + \xi + z\}^{1+\varepsilon}}, \qquad (5)$$

and taking into account, that $q^2 = sx$ we obtain (here $L = t/s$)

$$D(s,t) \Rightarrow \frac{1}{s^{3-2\varepsilon}} J,$$

$$J = \varepsilon \int_0^\infty d\xi \int_0^\infty d\zeta \int_0^\infty dx \int_0^\infty d\tau \frac{\xi\tau^{2\varepsilon-1}}{\{1+\tau+L\xi\}\{1+\tau+\zeta\}}$$

$$\times \frac{1}{\{x\zeta + (1+\tau)(x+\xi+\zeta)\}^\varepsilon} \int_0^\zeta dz \frac{1}{\{xz+x+z+\xi\}^{1+\varepsilon}}. \qquad (6)$$

This spectral representation is a convenient starting point for both analytical and numerical calculations. For instance, in a separate publication I will obtain one of the irreducible diagrams that appeared in [10], which corresponds to $L = -1$ as well as a much more cumbersome analytical result for arbitrary $L$ as a quadratic form of $Li_n$-functions.

*Conclusions.* In this paper we have considered an approach to the calculation of on-shell two-loop scalar double-box diagram. This approach is based on the following tools: (i) the Feynman parametric representation, (ii) the uniqueness method, (iii) Mellin-Barnes contour integrals. We considered only scalar diagrams (corresponding to massless $\varphi^3$-theory) because expressions occuring in realistic calculations can be reduced to such scalar integrals. We have obtained the analytical formula (6), that can be used for calculations in various concrete cases.

I am grateful to F.V.Tkachov for his interest in this work, useful discussions and help. The reseach was supported in part by the Russian Fund for Basic Reseach (grant 95-02-05794).